\documentstyle[12pt]{article}
\begin{document}
\baselineskip 3.9ex
\def\be{\begin{equation}}
\def\ee{\end{equation}}
\def\ba{\begin{array}{l}}
\def\ea{\end{array}}
\def\bea{\begin{eqnarray}}
\def\eea{\end{eqnarray}}
\def\no#1{{\tt   hep-th#1}}
\def\eq#1{(\ref{#1})}
\def\pgap{\vspace{1.5ex}}
\def\ggap{\vspace{10ex}}
\def\gap{\vspace{3ex}}
\def\del{\partial}
\def\o{{\cal O}}
\def\z{{\vec z}}
\def\re#1{{\bf #1}}
\def\av#1{{\langle  #1 \rangle}}
\def\S{{\cal S}}

\renewcommand\arraystretch{1.5}

\begin{flushright}
TIFR-TH-98/37\\
August 1998\\
hep-th/9808168
\end{flushright}
\begin{center}
\vspace{2 ex}
{\large{\bf Absorption and Hawking Radiation of
Minimal and Fixed Scalars, and AdS/CFT correspondence}}\\
\vspace{3 ex}
Justin R. David$^*$, Gautam Mandal$^\dagger$ and 
Spenta R. Wadia$^\ddagger$ \\
{\sl Department of Theoretical Physics} \\
{\sl Tata Institute of Fundamental Research,} \\
{\sl Homi Bhabha Road, Mumbai 400 005, INDIA.} \\
\vspace{10 ex}
\pretolerance=1000000
\bf ABSTRACT\\
\end{center}
\vspace{1 ex}

We present a complete derivation of absorption cross-section and
Hawking radiation of minimal and fixed scalars from the
Strominger-Vafa model of five-dimensional black hole, starting right
from the moduli space of the D1-D5 brane system. We determine the
precise coupling of this moduli space to bulk modes by using the
AdS/CFT correspondence. Our methods resolve a long-standing problem
regarding emission of fixed scalars. We calculate three-point
correlators of operators coupling to the minimal scalars from
supergravity and from SCFT, and show that both vanish. We make some
observations about how the AdS/SYM correspondence implies a close
relation between large $N$ equations of motion of $d$-dimensional
gauge theory and supergravity equations on $AdS_{d+1}$-type
backgrounds. We compare with the explicit nonlocal transform relating
1 and 2 dimensions in the context of $c=1$ matrix model.

\vfill
\hrule
\vspace{0.5 ex}
\leftline{$^*$ justin@theory.tifr.res.in}
\leftline{$^\dagger$ mandal@theory.tifr.res.in}
\leftline{$^\ddagger$ wadia@theory.tifr.res.in}
\clearpage

\section{Introduction}

In the past few years significant progress has been made in our
understanding of black hole physics in terms of string theoretic
models
\cite{Sus,Sen,StrVaf,CalMal,MalSus,DMW,DM,MalStr,CalGubKleTse,KraKle,MalD,MAL,JohKhurMye,BreLowMeyPeeStrVaf,Gub1,Gub2}.
Out of this, the derivation of black hole entropy from string theory,
based on a counting of BPS states \cite{StrVaf}, is an {\sl ab initio}
derivation. The discussion of dynamical issues like absorption and
Hawking radiation, however, is based on several plausible assumptions,
in particular (a) about the degrees of freedom of the D-brane system,
and (b) about how these couple to bulk quanta which appear as Hawking
radiation \cite{DMW,DM,MalStr}. It cannot be overemphasized that
without an {\sl ab initio} derivation of Hawking radiation, there will
be lingering doubts about any claimed explanation of black hole
thermodynamics and information loss within unitary quantum theory.  In
this paper we will present an {\sl ab initio} derivation of Hawking
radiation/absorption starting from the moduli space of low energy
degrees of freedom of the gauge theory describing D1-D5 system. We
will explicitly determine the gauge-invariant coupling of this moduli
space to minimal and fixed scalars and also construct in detail
microcanonical ensembles based on the moduli space leading to
gauge-invariant  $S$-matrix elements for absorption/emission. Since the
microscopic framework here is gauge theory, calculations based on it
are obviously unitary.

The low energy degrees of freedom of a large number of D1 and D5
branes in type IIB string theory compactified on $B_4$ ($B_4 = T^4$ or
$K3$) \cite{VafD,MalD,HasWad} in a nutshell are as follows. The
degrees of freedom of the D1-D5 system can be derived in one of two
ways. One is by regarding the D1 branes as instantons on the D5
branes, in which case the degrees of freedom are described in terms of
an instanton moduli space \cite{DouD}. This in turn is described in
terms of an ${\cal N}=(4,4)$ SCFT (superconformal field theory) based
on a resolution of the orbifold $(B_4)^{Q_1Q_5}/S(Q_1Q_5)$
\cite{VafD}. Here $S(p)$ denotes symmetric group of $p$ elements. The
second way is to describe the D1-D5 system in terms of a gauge theory
arising out of massless modes of various open strings that connect
these branes. The important component of this gauge theory are the
hypermultiplets which arise out of the open strings connecting D1 and
D5 branes \cite{MalD}. The low energy degrees of freedom of the system
have been explicitly solved (for $B_4 = T^4$) and correspond to the
hypermultiplet moduli space given by an ${\cal N}=(4,4)$ SCFT based on
$(B_4)^{Q_1Q_5}/[S(Q_1) \times S(Q_5)]$ \cite{HasWad}.  The two
representations in terms of instanton moduli space and the
hypermultiplet moduli space are conjectured to be equivalent
\cite{WitD}: it would certainly be worthwhile to understand the
equivalence in detail, in particular what the nonrenormalization
theorem for the hypermultiplet moduli space implies for the instanton
moduli space. This question has a bearing on the issue of
extrapolation from weak to strong coupling.

In what follows we will consider the SCFT based on (a resolution of)
the orbifold $(T^4)^{Q_1Q_5}/S(Q_1Q_5)$. (It is simple to extend our
results to the SCFT with the other quotient group.)  We will denote
the fields of the SCFT as $x^i_A(z,\bar z)$, $\psi_A^{a \alpha}(z)$
and $\bar \psi_A^{\dot a\dot \alpha}(\bar z)$. Here $i$ is the vector
index of $SO(4)^I$, the local Lorentz group of the 4-torus, and
$A=1,\ldots,Q_1Q_5$ labels $S(Q_1Q_5)$. Also, $a, \dot a$ denote
spinor labels of $SO(4)^I \equiv SU(2)^I \times SU(2)^I$, and $\alpha,
\dot \alpha$ denote spinor labels of the $R$-parity group $SO(4)^E
\equiv SU(2)^E \times SU(2)^E$ of ${\cal N}=(4,4)$ SCFT. The
superscript $E$ anticipates identification of $SO(4)^E$ later on in
supergravity with the isometry group of $S^3$ which is external to the
4-torus.  Besides the $x$'s and $\psi$'s we also have spin fields and
twist fields.

The plan of the paper is as follows. In Sec. 2 we describe the
absorption and emission of minimal scalars, specifically the traceless
symmetric deformation of the metric of the 4-torus. We first determine
the SCFT operator coupled to this field using the principle of
near-horizon symmetry underlying the AdS/CFT correspondence.  We
present a detailed discussion on how to determine the normalization of
the interaction Lagrangian.  We construct gauge-invariant density
matrices representing the black hole state and use the above coupling
to the bulk fields to calculate $S$-matrix for absorption and
emission. In Sec. 3 we present a calculation of 2- and 3-point
amplitudes of the SCFT operators from supergravity as dictated by the
quantitative version of the AdS/CFT conjecture. We also calculate
these amplitudes directly from SCFT. We show that the two-point
functions agree precisely for an appropriate choice of normalization
of the interaction Lagrangian, and that the three-point functions
vanish in either way of computing them. In Sec. 4 we discuss the
absorption/emission of fixed scalars and show how the existing
discrepancies between semiclassical and D-brane calculations disappear
once the correct coupling to SCFT operators is identified. In Sec. 5
we discuss intermediate scalars. In Sec. 6 we make some general
remarks about how the large $N$ equations of motion of gauge theories
are related to the equations of supergravity on $AdS$-type backgrounds
through the AdS/CFT correspondence. We also discuss how a similar
correspondence is effected in $c=1$ matrix model through a nonlocal
transform between 1 and 2 dimensions which we explicitly present.
Sec. 7 contains summary and concluding remarks.

\section{Minimal Scalars}

The massless spectrum of type IIB string theory compactified on $T^4$
\footnote{This 4-torus is not identical to the one appearing in the
SCFT described above, but this subtlety \cite{HasWad,GivKutSei} need
not concern us here.} has 25 scalars: the full spectrum is described
in Appendix B. Out of these scalars, five pick up masses when D1 and
D5 branes are introduced. The remaining twenty satisfy wave equations
appropriate for massless scalars minimally coupled to the metric of
the D-brane solution.  These are called minimal scalars.  We will
focus our attention on the familiar example of $h_{ij}$, the traceless
symmetric deformations of the 4-torus. 

A crucial ingredient in the D-brane method of computation of
absorption cross-section for these scalars or the rate of Hawking
radiation is the coupling of $h_{ij}$ to the D-branes.  This is 
given by a specific SCFT operator $\o_{ij}(\z)$ ($\z= (z, \bar z)$),
which {\sl couples} to the bulk mode $h_{ij}$ in the form of an
interaction 
\be
S_{\rm int}= \mu \int  d^2 \z [h_{ij}(\z) \o_{ij}(\z)]
\label{one}
\ee 
where $h_{ij}(\z)$ denotes the restriction of $h_{ij}$ to the
location of the SCFT, and $\mu$ is a number denoting the
strength of the coupling. Below we will discuss, given the operator
$h_{ij}$, first how to determine $\o_{ij}$ and later (below
equation \eq{free}) how to determine the constant $\mu$.

\subsection{Determination of the operator $\o_{ij}$:}

Method 1.  One way of determining the operator $\o_{ij}$ would be to
reanalyze the instanton moduli space or the hypermultiplet moduli
space with the metric of the $T^4$ deformed by $h_{ij}$. This method
is not very easy and we will not dwell on it any further.

\pgap

Method 2. A simpler but more elegant approach towards finding the
operator $\o_{ij}$ that couples to $h_{ij}$ is by appealing to {\sl
symmetries}. This method utilizes the dictionary between symmetries of
D-brane world-volume and those of spacetime. The steps are: (a) find
the symmetries $\S$ of the bulk, (b) find how (all or a part of) these
symmetries appear in D-brane world-volume and consequently how they
act on the variables of the SCFT, (c) find how $h_{ij}$ transforms
under $\S$, and (d) demand that $\o_{ij}$ should transform under the
{\sl same} representation of the symmetry group $\S$ when it acts on
the SCFT. The last step arises from the fact that $h_{ij}(\z)$ in
\eq{one} is a source for $\o_{ij}$. The hope is that this procedure
fixes $\o_{ij}$. It is clear that the normalization of $\o_{ij}$
would not be determined by these arguments. We will discuss
the determination of the normalization, or equivalently that of
the constant $\mu$ of equation \eq{one}, below \eq{free}. 

\pgap

It has long been recognized that the symmetries $\S'= SO(4)^I\times
SO(4)^E$ of the bulk theory (local Lorentz rotation of the 4-torus and
rotation in the transverse space) appear naturally in the SCFT of the
D-brane world volume as well. The $SO(4)^I$ part is obvious; $SO(4)^E$
appears as the R-parity group (see \cite{WitD}, {\sl e.g.}).  Let us
now apply steps (c) and (d) above in the context of this symmetry
group $\S'$.

The field $h_{ij}$ (symmetric, traceless) transforms as $(\bf 3, \bf
3)$ under $SO(4)^I\equiv SU(2)^I \times SU(2)^I$ and as $(\bf 1, \bf
1)$ under $SO(4)^E \equiv SU(2)^E \times SU(2)^E$. 

Now there are at least three possible SCFT operators which belong to
the above representation of $\S'$:
\bea
\o_{ij} &=& \del x^i_A \bar \del x^j_A 
\nonumber \\
\o'_{ij} &=& \psi^{\alpha}_{aA}(z) \sigma_i^{a \dot a}
\bar\psi^{\dot \alpha}_{\dot a B}(\bar z) 
\psi_{\alpha,bA}(z) \sigma_j^{b \dot b}
\bar\psi_{\dot \alpha,\dot b B}(\bar z)  
\nonumber \\
\o''_{ij} &=& \psi^{\alpha}_{aA}(z) \sigma_i^{a \dot a}
\bar\psi^{\dot \alpha}_{\dot a A}(\bar z) 
\psi_{\alpha,bB}(z) \sigma_j^{b \dot b}
\bar\psi_{\dot \alpha,\dot b B}(\bar z)  
\label{two}
\eea
The spinor labels are raised/lowered above using the
$\epsilon^{\alpha\beta},\epsilon_{\alpha\beta}$ symbol. The
$\sigma_i$'s denote the matrices : $(1, i\vec \tau)$.  The last two
operators differ only in the way the $S(Q_1Q_5)$ labels are
contracted. All the three operators should be regarded as symmetric
(in $(i,j)$) and traceless.

The complete list of operators with the same transformation property
under $\S'$ contains, in addition, those obtained by multiplying any
of the above by singlets. These would necessarily be irrelevant
operators, but cannot be ruled out purely by the above symmetries.

It might seem `obvious' that the operator $\o_{ij}$ should be the
right one to couple to the bulk field $h_{ij}$. However, the simplest
guesses can sometimes lead to wrong answers, as we will see
later for fixed scalars, where it will turn out that the operator
$\del x^i_A \bar \del x^i_A$ is far from being the right one to couple
to $h_{ii}$ (trace). We proceed, therefore, to find out the right
operator, by sticking to the principle stated in Method 2 above.

\subsection{Incorporation of Near-horizon Symmetry}

It has been conjectured recently \cite{Malads,GubKlePol,Witads} that
if one takes the large $gQ$ ($Q=Q_1, Q_5$) limit, then a powerful
correspondence can be built between the physics of the bulk and the
physics of the boundary.  This has many qualitative and quantitative
consequences. For the limited purpose of identifying the SCFT
operator, it is enough to use only the most basic part of the
conjecture which says that going to the large $gQ$ limit leads to an
enhancement of the symmetry. Since the `proof' of this part is
obvious, we will accept results based on this as derived from first
principles.

The discussion below has overlap with a number of recent works
\cite{Mar,MalStr98,Lar,deB,GivKutSei}. These papers, especially the
work of de Boer \cite{deB}, provide the background for many results of
the present paper.  We have found it useful to carry out a detailed
extension of de Boer's method to the low supergravity modes,
especially in the case of $T^4$, which we have included in Appendix
B (see also \cite{Lar}). 

In the large $gQ$ for the present system the symmetry group $\S'$ is
enhanced to $\S=SO(4)^I\times SO(4)^E \times SU(1,1|2) \times
SU(1,1|2)$.

{}From the spacetime point of view, this happens because in this limit
the spacetime geometry is effectively the {\sl near-horizon geometry}
of a D1-D5 system (wrapped on $T^4 \times S^1$) which is $AdS_3 \times
S^3 \times T^4$ (with $x^5$ periodic) with appropriate values of the
RR two-form field $B'$ (see Appendix A for details)\footnote{The
near-horizon geometry of the finite-area black hole actually involves
a BTZ black hole ($\times S^3 \times T^4$) which corresponds to the
Ramond sector of the SCFT, whereas $AdS_3$ corresponds to the NS
sector.  However, the local operators of interest here can be shown to
have the same symmetry properties irrespective of whether they belong
to the Ramond or the NS sector (\cite{MalStr98,GivKutSei}, see also
remarks at the end of Appendix B).  The ground state of the Ramond
sector is degenerate, as against that of the NS sector; this
degeneracy would be reflected in our construction of the black hole
state \eq{eq.state} --- however, this would not affect the $S$-matrix
relevant for absorption and emission.  We consider below the NS
sector, corresponding to $AdS_3$, but for the specific purposes of the
paper the discussion is equally applicable to the Ramond sector. }.
The factor $SO(4)^I$ acts as before. The group $SO(4)^E$ corresponds
in the new picture to the isometry group of $S^3$. The bosonic part of
$SU(1,1|2)\times SU(1,1|2)$ arises as the isometry group of $AdS_3$
(which is the $SL(2,R)$ group manifold).  The $SU(2)$ part which
transform the fermions among themselves, and the off-diagonal
supersymmetry transformations, are a consequence of ${\cal N} = (4,4)$
supersymmetry of this compactification.  On the SCFT side, the $SO(4)$
groups have actions as before.  The $SU(1,1|2)$ is identified with the
subgroup of the superconformal algebra generated by $L_{\pm 1,0},
G^{a\alpha}_{\pm 1/2}$ (the other $SU(1,1|2)$ involves $\bar L, \bar
G$).
 
Let us now apply steps (c) and (d) of Method 2 to this enhanced
symmetry group $\S$. How does $h_{ij}$ transform under $SU(1,1|2)$?
{}From the fact that $h_{ik}$ transforms as $({\bf 1}, {\bf 1})$ of
$SO(4)^E$ we can deduce that its various KK (Kaluza-Klein) modes
$h_{ik}^{j,j'}$ will obey the restriction $j=j'$ (see Appendix B for
details). If we restrict ourselves for the present to $s$-waves we
have $j=j'=0$. Now since the $h_{ij}$ is a massless (minimal) scalar,
it corresponds to $(L_0,\bar L_0)=(1,1)$ where by $L$'s here we mean
$SU(1,1)$ generators in the bulk ({\sl cf.}, \cite{MalStr98}).  Since
$h_{ij}$ creates single-particle excitations, let us classify it as a
short multiplet (more on this later). Looking at the list (Appendix B)
of short multiplets, we find that there is only one short multiplet of
$SU(1,1|2)$ which contains the field $j=0, L_0=1$: viz.  $({\bf
2},{\bf 2})_S$ of $SU(1,1|2) \times SU(1,1|2)$ (see Appendix B for
notation).  It is important to note that the ($j=0, L_0=1$) field
occurs as the `top' component (killed by $G_{-1/2}$, and not by
$G_{1/2}$) of that supermultiplet.

According to step (d) we now look for a SCFT operator $\o_{ij}$ which
is the top component of a $(\re{2},\re{2})_S$ short supermultiplet of
$SU(1,1|2)\times SU(1,1|2)$ and also has $(h,\bar h)=(1,1)$. 
We need to find which of
the operators $\o_{ij}, \o'_{ij}$ and $\o''_{ij}$ has this property.
Note that operators obtained by multiplying with nontrivial $\S'$
singlets does not have $(h,\bar h)=(1,1)$. Now, it is easy to see that
only $\o_{ij}$ is killed by the `raising' operators $G_{-1/2}$.
Coupled with the fact that it has $j=0, L_0=1$ (and similar equation
for the antiholomorphic sector), it matches the transformation
property of $h_{ij}$ completely. The other two operators are killed
by the lowering operators $G_{1/2}$. So they are `bottom' components
of a supermultiplet. Now, bottom components of all $SU(1,1|2)$
supermultiplets have $j=L_0$. Since both $\o'_{ij}$ and $\o''_{ij}$
have $j=0 < L_0=1$, they cannot be the right operators to couple
to $h_{ij}$.

{\sl Hence, we find that $\o_{ij}$ is indeed the right operator to
couple to $h_{ij}$ in Equation \eq{one}.}

This choice was independently arrived at in \cite{HasWad} from their
analysis of the hypermultiplet moduli space. The variable $x^i_A$ was
denoted there as $y^i_{aa'}$ where $a,a'$ are $S(Q_1)$ and $S(Q_5)$
indices respectively. This operator also appears in \cite{Mar}.

Note that this derivation assumes that $h_{ij}$ (like other fields in
the supergravity spectrum, Appendix B) should belong to short
multiplets. This assumption is vindicated by the complete accounting
of {\sl all} KK modes on $S^3$ in terms of short multiplets, as we
show on Appendix B (see \cite{deB} for a more detailed discussion
of this issue).

\subsection{ Absorption and Hawking Radiation:}

With the above result in hand, we can now use \eq{one} to perform a
D-brane computation of absorption cross-section and Hawking radiation
for minimal scalars.

Instead of detailing the entire computation we will emphasize
the essential conceptual differences from  earlier works \cite{DMW,DM}. 
{}From the above discussion, the interaction Lagrangian is
\be
\label{interaction} 
S_{\rm int} = \mu T_{\rm eff} \int d^2 z\; 
\left[ h_{ij} \del_z x^i_A \del_{\bar z} x^j_A \right]
\ee
The effective string tension $T_{\rm eff}$ of the conformal
field theory , which also appears in the free part of the action 
\be
\label{free}
S_0 = T_{\rm eff} \int d^2 z\; \left[\del_z
x^i_A \del_{\bar z} x_{i,A} + {\rm fermions} \right]
\ee
has been discussed in \cite{CalGubKleTse,MAL,HasWad}. The specific
value of this is not important for the calculation of the $S$-matrix
for absorption or emission, since the factor cancels in the $S$-matrix
between the interaction Lagrangian and the external leg factors. 
However, the value of $\mu$ {\sl is important} to determine
since the absorption crosssection and Hawking radiation rates
calculated from the SCFT depend on it.
We will argue below that $\mu=1$. 

\pgap

\noindent\underbar{Determination of the normalization constant
$\mu$ in $S_{\rm int}$:} 

\pgap

A direct string theory computation, as in Method 1 mentioned in the
context of determining the operator $\o_{ij}$, would of course provide
the constant $\mu$ as well (albeit at weak coupling perhaps).  This
would be analogous to fixing the normalization of the
Dirac-Born-Infeld action for a single D-brane by comparing with a
one-loop open string diagram \cite{Pol}. However, for a large number,
and more than one type, of D-branes it is a very difficult proposition
and we will not attempt to pursue it here. Fortunately, as in Method 2
for determining the operator $\o_{ij}$, the AdS/CFT correspondence
helps us determine the value of $\mu$ as well. For the latter,
however, we need to use the more quantitative version
\cite{Witads,GubKlePol} of the conjecture.  In Section 3, we will see
that for this quantitative conjecture to be true for the two-point
function (which can be calculated independently from the ${\cal N}=4$
SCFT and from supergravity) we need $\mu=1$.

We will see below that the above normalization leads to precise
equality between the absorption cross-sections (and consequently
Hawking radiation rates) computed from the moduli space of the D1-D5
system and from semiclassical gravity. This method of fixing the
normalization can perhaps be criticized on the ground that it borrows
from supergravity and does not rely entirely on the SCFT.  However, we
would like to emphasize two things:\\ 
(a) We have fixed $\mu=1$ by
comparing with supergravity around AdS$_3$ background which does 
{\sl not} have a black hole.  On the other hand, 
the supergravity calculation of absorption
cross-section and Hawking flux is performed around a {\sl black hole
background} represented in the near-horizon limit by the BTZ black
hole. From the viewpoint of semiclassical gravity  these two 
backgrounds are rather different. The fact that normalizing
$\mu$ with respect to the former background leads to the
correctly normalized absorption cross-section around the
black hole background is a rather remarkable prediction.\\
(b)  Similar issues are involved in fixing the coupling
constant between the electron and the electromagnetic field in the
semiclassical theory of radiation in terms of the physical electric
charge, and in similarly fixing the gravitational coupling of extended
objects in terms of Newton's constant.  These issues too are decided
by comparing two-point functions of currents with Coulomb's or
Newton's laws respectively. In the present case the quantitative
version of the AdS/CFT conjecture \cite{Witads,GubKlePol} provides the
counterpart of Newton's law or Coulomb's law at strong coupling.
Without this the best result one can achieve is that the Hawking
radiation rates computed from D1-D5 branes and from semiclassical
gravity are {\sl proportional}.

We should remark that fixing the normalization by the use of
Dirac-Born-Infeld action, as has been done previously, is not
satisfactory since the DBI action is meant for single D-branes and
extending it to a system of multiple D1-D5 branes does not always give
the right results as we will see in the section on fixed scalars. The
method of equivalence principle to fix the normalization is not very
general and cannot be applied to the case of non-minimal scalars, for
example. 

With these comments, we now go back to the calculation of the
$S$-matrix.

The black hole is represented by a density matrix
\be 
\rho = \frac{1}{\Omega} \sum_{\{i\}} | i \rangle \langle i |
\label{density}
\ee 
This is the same as Equation (9) of \cite{DMW}. However, the
states $| i \rangle$ now represent gauge-invariant states (invariant
under $S(Q_1Q_5)$) from all possible twisted sectors of the 
orbifold SCFT. 

The explicit formula for these states $| i \rangle$ for an arbitrary
twisted sector is somewhat involved. Since the maximally twisted
sector, defined by the permutation element
\be
g:x^i_A\to x^i_{A+1},
\label{deftwist}
\ee 
has dominant contribution ({\sl cf.}  \cite{HasWad}) to the
density matrix \eq{density}, let us write out the gauge-invariant
states $| i \rangle$ for this sector. The variables $x^i_A(z, \bar z)$
belonging to this sector satisfy: 
\be 
x^i_A(\sigma+2\pi, t) = g(x^i_A)(\sigma,t) \equiv x^i_{A+1}(\sigma,t)
\label{twist1}
\ee
In the above we define $A+1\equiv1$ when $A=Q_1Q_5$.
Similar equations hold for the fermions. 

Let us define a periodic variable $\tilde x^i(\sigma,t)$ on a larger
circle $\sigma\in [0, 2\pi Q_1Q_5)$ \cite{MooDVV,DVV} by
\be
\tilde x^i(\sigma + 2\pi(A-1),t) \equiv x^i_A(\sigma,t),
\sigma\in [0,2\pi )
\ee
which will have a normal mode expansion:
\be
\tilde x^i(\sigma,t)= (4\pi T_{\rm eff})^{-1/2} 
\sum_{n>0} \left[\left(
 \frac{a^i_n}{\sqrt n} e^{i n(-t+\sigma)/Q_1Q_5} + 
\frac{\tilde a^i_n}{\sqrt n} e^{i n(-t-\sigma)/Q_1Q_5} \right)
+ {\rm h.c.} \right]
\ee
The twist \eq{deftwist} acts on these oscillators as
\bea
\label{twist2}
g:a^i_n &\to& a^i_n e^{2\pi i n/Q_1Q_5} \nonumber\\
g:\tilde a^i_n &\to& \tilde a^i_n e^{-2\pi i n/Q_1Q_5}
\nonumber\\
\eea
The states $|i\rangle$ are now defined as 
\be
\label{eq.state}
|i\rangle = \prod_{n=1}^{\infty}\prod_{i} C(n,i) 
(a^{i\dagger}_n )^{N_{L,n}^i} (\tilde a^{i\dagger}_n )^{N_{R,n}^i}
|0 \rangle
\ee 
where $C(n,i)$ are normalization constants ensuring unit norm of
the state ({\sl cf.} Equation (3) of \cite{DMW}, which used some given
polarization index $i$). $|0 \rangle$ represents the NS ground
state. As explained in the footnote on page 6, the present discussion
is also valid in the Ramond sector, in which case the ground state
will have an additional spinor index but will not affect the
$S$-matrix. We have also suppressed the fermion creation operators
which also do  not affect the $S$-matrix. 

It is clear that the creation operators create KK momentum along $S^1$
(parametrized by $x^5$). The total left (right) moving KK momentum of
\eq{eq.state} (in units of $1/\tilde R, \tilde R \equiv Q_1Q_5 R_5$,
$R_5$ being the radius of the $S^1$) is $N_L$ ($N_R$),
where
\be
\label{totaln}
N_L = \sum_{n,i} n N^i_{L,n}, \quad N_R = \sum_{n,i} n N^i_{R,n} 
\ee
{}From \eq{twist2} and \eq{eq.state}, we see that
\be
g:|i\rangle \to \exp[\frac{2\pi i}{Q_1Q_5} (N_L-N_R) |i\rangle
\label{twist3}
\ee
Now, the total KK momentum carried by $|i\rangle$
is $p_5= (N_L - N_R)/(Q_1 Q_5 R_5)$. Quantization of the KK charge
requires that $p_5 = \hbox{integer}/R_5$, which implies that
\be
\label{twist4}
(N_L - N_R)/(Q_1 Q_5) = \hbox{integer}
\ee
Thus, using \eq{twist3} and the above equation, we find 
the states $|i \rangle$ representing microstates of the
black hole to be  gauge invariant\footnote{In \eq{totaln} we
have written down only the bosonic contribution to the 
KK momentum. If we wrote the total contribution of bosons
and fermions, then taking into account an identical 
gauge transformation property for the fermionic oscillators
as in \eq{twist2} we would arrive at the same 
conclusion as above, {\sl viz}, that the state $|i\rangle$ 
is gauge invariant.}.

The rest of the calculation now follows formally along the lines of
\cite{DMW,DM} and the final results obtained are the same, thus
establishing the agreement between the D-brane calculation and the
semiclassical calculation.

The present discussion provides, in our perception for the first time,
a complete derivation of absorption and Hawking radiation from the
five-dimensional black hole.

\section{Two- and Three-point Functions of Minimal Scalars}

In this section we will discuss the more quantitative version of the
AdS/CFT conjecture \cite{GubKlePol,Witads} to compare 2- and 3-point
correlation functions of $\o_{ij}$ from supergravity and from SCFT
respectively.

The relation between the correlators  are as
follows. Let the supergravity Lagrangian be
\bea 
\label{lag}
L &=& \int d^3 x_1 d^3 x_2 b_{ij,i'j'}(x_1,x_2) h_{ij}(x_1) 
h_{i'j'}(x_2)\nonumber\\
 &+& \int d^3 x_1 d^3 x_2 d^3
x_3 c_{ij,i'j',i''j''}(x_1, x_2, x_3) h_{ij}(x_1) h_{i'j'}(x_2)
h_{i''j''}(x_3) + \ldots\nonumber\\ 
\eea 
where we have only exhibited terms quadratic and cubic in the 
$h_{ij}$'s. The coefficient $b$ determines the propagator and
the coefficient $c$ is the tree-level 3-point vertex in supergravity. 
$b$ and $c$  can be read out from Appendix A.

The 2-and 3-point functions of the $\o_{ij}$'s (at large $gQ$)
are then given by \cite{GubKlePol,Witads}, assuming an $S_{\rm int}$
given by \eq{interaction}
\bea
\label{2pt}
& \langle \o_{ij}(z_1) \o_{i'j'}(z_2)\rangle
\\    \nonumber 
& = 2(\mu T_{\rm eff})^{-2}
\int d^3 x_1 d^3 x_2 \left[ b_{ij,i'j'}(x_1, x_2) 
K(x_1|z_1) K(x_2|z_2)\right],
\eea
\bea
\label{3pt}
& \langle \o_{ij}(z_1) \o_{i'j'}(z_2) \o_{i''j''}(z_3) \rangle
\\    \nonumber 
& = 3! (\mu T_{\rm eff})^{-3} 
\int d^3 x_1 d^3 x_2 d^3 x_3
\left[ c_{ij,i'j',i''j''}(x_1, x_2, x_3) 
K(x_1|z_1) K(x_2|z_2) K(x_3|z_3) \right]
\eea
where $K$ is the boundary-to-bulk Green's function for massless
scalars \cite{Witads}
\be
\label{green}
K(x|z)= \frac{1}{\pi} \left[ \frac{x_0}{(x_0^2 + (|z_x - 
z|^2)}\right]^2
\ee
We use complex $z$ for coordinates of the CFT, and
$x = (x_0, z_x)$ for the (Poincare coordinates) of bulk theory.

\noindent\underbar{Two-point function:}

The right hand side of \eq{2pt} can be evaluated  using equation (17)  
in \cite{FreMatMatRas}, with $\eta= Q_1 Q_5/(8 \pi)$ in our case,
where we have been  careful to convert to complex coordinates 
at the boundary. We find that
\be
\langle \o_{ij}(z) \o_{i'j'}(w) \rangle =
(\mu T_{\rm eff})^{-2}\delta_{ii'}\delta_{jj'} 
\frac{Q_1 Q_5}{16 \pi^2}\frac{1}
{|z - w|^4}
\ee  
This is exactly the value of the two-point function obtained
from the SCFT described by the free Lagrangian \eq{free} provided
we put $\mu=1$. It is remarkable that even at strong coupling
the two-point function of $\o_{ij}$ can be computed 
from the free Lagrangian
\eq{free}.  This is consistent with the nonrenormalization 
theorems involving the ${\cal N}=4$ SCFT.

The choice $\mu=1$ ensures that the perturbation \eq{interaction}
of \eq{free} is consistent with the perturbation implied in  
\eq{changemetric}. We have already remarked in Section 2.3
that this choice leads to precise equality between absorption
cross-sections (consequently Hawking radiation rates) calculated from
semiclassical gravity and from the D1-D5 branes. 

\noindent\underline{Three-point function:}

Before proceeding to evaluate \eq{3pt}, let us pause to see what a
tree-level CFT calculation of the three-point correlator of
the $\o_{ij}$'s gives. It is straightforward to see that 
the three-point function vanishes:
\be
\label{cft}
\langle \o_{ij}(\z_1) \o_{i'j'}(\z_2) \o_{i''j''}(\z_3) \rangle=0
\ee
The reason is simply that the correlator splits into a holomorphic
and an antiholomorphic factor, and in each of them there are
an odd number of $x$'s. Therefore each factor vanishes (ignoring
possible contact terms throughout). Interestingly, the 
three-point function of both $\o'_{ij}$ and  $\o''_{ij}$
are nonvanishing.

The vertex factor $c$, as seen from Appendix A, is clearly
non-vanishing. 
So \eq{cft} seems to be at variance with the fact that the vertex
factor $c$ in \eq{3pt} is nonzero. Before going to ascribe the
difference to strong and weak coupling, let us evaluate the r.h.s. of
\eq{3pt}.

Using the list of integrals in \cite{FreMatMatRas} (eqs. 19,20,25,29)
we find, rather remarkably, that \eq{3pt} gives a zero answer too!
Note that there are two surprises leading to this answer: (a) the
vanishing of the integral in \eq{3pt}, as just mentioned, and (b) the
vanishing of the coefficient of the cubic term coupled to RR
backgrounds; if this coefficient were not to vanish, the corresponding
integral in \eq{3pt} would have been different from the one coming
from terms coupled to the metric, and would have been nonzero.

\pgap

Thus, the three-point function calculated from the CFT
(ostensibly at weak coupling) and that calculated from the
supergravity agree, and both vanish!

\section{Fixed Scalars}

Out of the 25 scalars mentioned earlier which form part of the
spectrum of IIB supergravity on $T^4$, five become massive when
further compactified on $AdS_3 \times S^3$. There is an important
additional scalar field which appears after this compactification:
$h_{55}$. Our notation for coordinates here is as follows: $AdS_3:
(x^0, x^5, r), S^3:(\chi, \theta, \phi); T^4:(x^6, x^7, x^8,
x^9)$. $r, \chi,\theta,\phi$ are spherical polar coordinates for the
directions $x^1,x^2,x^3,x^4$. In terms of the D-brane wrappings, the
D5 branes are wrapped along the directions 5-9 and D1 branes are
wrapped along direction $5$. The field $h_{55}$ is scalar in the sense
that it is a scalar under the local Lorentz group $SO(3)$ of $S^3$.

In what follows we will specifically consider the three scalars
$\phi_{10}, h_{ii}$ and $h_{55}$. The equations of motion of these
fields in supergravity are coupled and have been discussed in detail
in the literature \cite{CalGubKleTse,KraKle}. It turns out that a
linear combination of $h_{ii}$ and $\phi_{10}$ remains massless; it is
part of the twenty massless (minimal) scalars previously discussed.
Two other linear combinations $\lambda$ and $\nu$ satisfy coupled
differential equations. These are examples of what are called `fixed
scalars'.

Understanding the absorption and emission properties of fixed scalars
is an important problem, because the D-brane computation and
semiclassical black hole calculation of these properties appear to be
at variance \cite{KraKle}.  The discrepancy essentially originates
from the `expected' couplings of $\lambda$ and $\nu$ to SCFT operators
with $(h, \bar h)=(1,3)$ and $(3,1)$.  These SCFT operators lead to
qualitatively different greybody factors from what the fixed scalars
exhibit semiclassically\footnote{For earlier attempts at understanding
this difference see \cite{LeeMyuKim}. }. The semiclassical greybody
factors are in agreement with D-brane computations if the couplings
were only to (2,2) operators. The precise agreement depends on the
normalization of this coupling which is given in terms of the
effective string tension, discussed at length in
\cite{CalGubKleTse,HasWad}\footnote{The normalization could also be
`determined' by demanding \cite{Teo} that the two-point functions in
the bulk and at the boundary agree in the sense of
\cite{Witads,GubKlePol}.}.

The coupling to $(1,3)$ and $(3,1)$ operators is guessed from
qualitative reasoning based on the Dirac-Born-Infeld action. Since we
now have a method of {\sl deducing} the couplings $\int \phi \o$ based
on near-horizon symmetries, let us use it to the case of the fixed
scalars.

The steps are similar to the case of the minimal scalars, so we will
be brief. It is obvious that the fixed scalars transform as $({\bf 1},
{\bf 1})$ of $SO(4)^E$ (and of $SO(4)^I$ as well, although that will
not play any role). Furthermore, the equations of motion for $\lambda$
and $\nu$ decouple in the near-horizon limit (this point has not been
emphasized much in the literature), and each corresponds to a massive
scalar field with mass $m^2 = 8/R^2$ (where $R$ is the radius of
curvature of $AdS_3$ defined in Appendix A). By scale invariance of
the interaction term $\int \lambda \o$ we deduce that $h + \bar h = 4$
(similarly for $\nu$)\footnote{If we had assumed, as in
\cite{MalStr98}, that these fields are Virasoro primaries, then we
could have read off the spins as well; however, we can assume this
generally only for `bottom' components.}. This is of course compatible
with $(h, \bar h)= (2,2), (1,3), (3,1), (0,4)$ and $(4,0)$ (note that
for $h+ \bar h=0$ we automatically have $h=\bar h=0$). As before, we
now see which of these values, together with $(j, \bar j)= (0,0)$,
occur, in short multiplets of $SU(1,1|2)$. We find that $(h,\bar h)=
(2,2)$ is the only choice.

We also find that the fixed scalars belong to the
short multiplet $( \re 3 , \re 3 )_S$ of
$SU(1,1|2) \times SU(1,1|2)$. This,
together with the fact that $(h,\bar h)= (2,2)$ occurs as the `top'
component of the supermultiplet, leads to only two SCFT operators
\bea
\o_1 &=& \del x^i_A \del x^i_A \bar \del x^j_B \bar \del x^j_B
\nonumber \\
\o_2 &=& \del x^i_A \del x^j_A \bar \del x^i_B \bar \del x^j_B
\label{fixed}
\eea 
corresponding to the two bulk fields $\lambda$ and $\nu$.  Which
specific linear combinations of these couple to the two fields
respectively, remains undetermined at this stage, but the D-brane
calculation for absorption/emission using either leads to the same
result\footnote{The appearance of the $S(Q_1Q_5)$ indices $A,B$
introduces considerable subtlety into the D-brane calculations.  As in
the case of minimal scalars, the dominant contribution comes from the
maximally twisted sector; so one can again introduce the extended
variable $\tilde x^i(\sigma)$. However, both $\o_1$ and $\o_2$ contain
products of operators which are generally at two different points
$\sigma, \sigma'$ of the circle. The calculation of the $S$-matrix can
still be carried out and it can be shown that the absorption
cross-section or rate of Hawking radiation is not changed by this to
leading order in frequency $\omega$ of the bulk quantum.}.  This
accords with the semiclassical calculations since $\lambda$ and $\nu$
satisfy identical differential equation, leading to the same
absorption/emission properties. We emphasize that in this analysis
too, we have assumed that the fixed scalars should form short
multiplets. This assumption is amply justified in Appendix B, where
all the KK modes on $S^3$ are correctly classified as a result of this
assumption ({\sl cf.}  \cite{deB}).

\gap

{\sl In summary, since the $(1,3)$ and $(3,1)$ operators are ruled out
by our analysis, the discrepancy between the D-brane calculation and
the semiclassical calculation of absorption and emission rates
disappears. It is important to note here that couplings guessed from
reasoning based on Dirac-Born-Infeld action turn out to be incorrect.}

\section{Intermediate Scalars}

We just make the remark that the classification presented in Appendix
B correctly account for all sixteen intermediate scalars, and predict
that they should couple to SCFT operators with $(h,\bar h) =(1,2)$
belonging to the short multiplet $({\bf 2}, {\bf 3})_S$ or operators
with $(h,\bar h) =(2,1)$ belonging to the short multiplet $({\bf 3},
{\bf 2})_S$. This agrees with the `phenomenological' prediction made
earlier in the literature \cite{KleRajTse}.

\section{Large $N$ classical equations of motion of gauge theories}

In previous sections, the superconformal field theory arose from a
large $N$ gauge theory (in either description of the moduli space).
The aspect of large $N$ that was used there was that in the large $N$
(more precisely large $gQ$) limit, the symmetries of the gauge theory
and those of the supergravity solution could be identified. The
precise role of large $N$ in the gauge theory as such was not used.
In this section we make some remarks concerning this issue.  In
particular, we discuss elements of large $N$ classical dynamics of
gauge theories are encoded in AdS spacetimes through the AdS/SYM
correspondence and also discuss how a similar correspondence appears
in $c=1$ matrix model.

One of the most important realizations that came out of the study of
the large N limit of field theories (including gauge theories) is the
fact that the large N limit can be formulated as a systematic
semiclassical expansion in 1/N. The theory is formulated in terms of
appropriate operators which satisfy the factorization condition at
large N: $\av{X^2}=\av{X}^2 + o(1/N)$ \cite{largeN}. For example in
gauge theories such operators are $(1/N\,) {\rm tr}\,W(C)^{n}$
($W(C)$ is the Wilson loop operator).  In the
c=1 matrix model we have the collective fields $(1/N\,) {\rm tr}\,
\exp(ikM)$.  These quantities (or appropriate variables constructed
out of them, such as the density variable $\rho(\lambda,t)$ or fermion
bilinear $\psi(x,t)\psi^\dagger(y,t)$ of $c=1$ models \cite{DMWc1} )
satisfy a set of classical equations of motion valid at
$N=\infty$. Let us denote such equations, generically as
\be
\label{zero}
{\cal F}(\Phi) = 0
\ee
The classical solution $\Phi_0$ of these equations of motion is
called the `master field'. Fluctuations around this 
solution, defined by
\be
\Phi= \Phi_0 + \frac{1}{\sqrt N}\delta \Phi
\ee
satisfy linear equations (to $o(1)$)
\be
\label{masterone}
\left. \frac{\del {\cal F}}{\del \Phi}\right|_0  \delta \Phi=0 
\ee 
giving the spectrum of
the theory at large $N$. The $o(1/N)$ and higher terms involve
$\del^2{\cal F}/(\del \Phi)^2|_0$ etc.  represent interaction. A
simple example of such a procedure can be found in \cite{DMWqcd} where
2-dimensional QCD is solved in terms of fermion bilinears whose
`master field' is presented explicitly.

We will illustrate how the AdS/SYM correspondence at large
$\lambda=gN$ essentially determines the various coefficients
in the Taylor series expansion of ${\cal F}$ around $\Phi_0$
except the solution $\Phi_0$ itself. Let us consider the
example \cite{Witads2} of the confining phase of the d=3 YM theory at
large N and large $\lambda$. This is {\sl dual} to the AdS
Schwarzschild black hole ($X_{2}$ in the notation of
\cite{Witads2}). An analysis of the solution of the scalar wave
equation indicates an asymptotic solution given by
\be
\label{wita}
\phi(\rho,x) \sim e^{ik.x}/\rho^4, \; \rho\to\infty
\ee
where $x$ denotes coordinates on the boundary ($\rho\to\infty$),
assumed Euclidean. An analysis of the full solutions $\phi(\rho,x)$
shows that normalizable solutions occur only at discrete
values $k^2= -m_n^2<0$. Any of these solutions $\phi_n(\rho,x)$,
therefore, leads to a wave on the boundary satisfying the equation
\be
\label{witb}
(-\del_t^2 + \del_{\bf x}^2 + m_n^2) \psi_n (x) = 0
\ee
where we have Wick rotated the equation to Lorentzian signature to
emphasize that this corresponds to a physical particle. Indeed,
this represents a scalar glueball of mass $m_n$. 

Equation \eq{witb} can be regarded as the equation for the expectation
value of a small Wilson loop in 2-dimensions. In the light of the
remarks made in the beginning of this section, this equation is the
linear equation \eq{masterone} for fluctuations around the large-$N$
classical solution.  In terms of a Lagrangian formulation, this tells
us about the quadratic fluctuation operator around the classical
solution, for instance that the eigenvalues of this operator are
discrete and calculable from supergravity \cite{Oog,KocJevMihNun}.
 More detailed information about this operator can be
obtained by computing the two-point function of various ${\rm tr}\,
F^{2n}$ operators from the AdS supergravity and looking at their
spectral distributions.

If we carry on to compute the various $n$-point correlations from
AdS supergravity, we can reconstruct the various orders of the
large N equation around the classical solution. It is interesting
to note that the classical solution itself cannot be obtained
by this method. It is tempting to think that the knowledge
of this solution must be tied to the choice of the specific
solution of supergravity.

\pgap

\underline{\sl $c=1$ matrix model:}\footnote{This
section has been developed in collaboration with Avinash Dhar.}

\pgap

It is clear that in the above example the knowledge of the spectrum in
the gauge theory is tied to the differential equation in the AdS
background. Alternatively it is related to the bulk-to-boundary
Green's function ({\sl cf.} \eq{green}) in this particular background.
Such a Green's function relating a $d$-dimensional physics to
a $d+1$ dimensional physics is already familiar from our study
of \cite{c1}. We refer the reader for detail to these references,
but the essential point is this: 

The $c=1$ matrix model is a theory of one-dimensional $N\times N$
matrices $M_{ij}(t)$ (much like a `gauge theory' in $d=1$ would
be). Variables of this theory at large $N$ can be related by nonlinear
nonlocal transforms to `spacetime' fields living in 1+1 dimensions
(related to the `tachyon' of two-dimensional string theory). The
specific transform is given in the form 
\bea 
\label{transform}
{\cal T}(x,t) &=& \sum_{p,q} G_{1,p,q} \delta u_{p,q}(t) + \ldots
\nonumber\\ 
\delta u_{p,q}(t) &=& u_{p,q}(t) - u_{0,p,q}(t)
\nonumber\\ 
u_{p,q}(t) &=& \int d\lambda e^{-ip\lambda}
\psi^\dagger(q-\lambda/2,t)\psi(q+\lambda/2,t) 
\eea 
Here
$\psi(\lambda,t)$ denotes the fermion field of the $c=1$ matrix model
\cite{c1}. The ellipsis denotes non-linear terms in the transform
({\sl cf.} Eqn. (3.4) of \cite{DMWdiscrete}).  The exact details of
these equations are given in \cite{DMWdiscrete,Avireview}. Just like
in the gauge theory example discussed above, the equations of motion
for the master field $u_{p,q}(t)$ (called $u(p,q,t)$ in the references
just mentioned) in the 1-dimensional theory get related to the
equations for the 1+1 dimensional fields. Also like above, the
interactions of the one-dimensional fields are related to those of the
two-dimensional fields ${\cal T}(x,t)$ through this transform.

This indeed represents a holographic realization of $c=1$ matrix
model\footnote{While this paper was being completed, we received
\cite{Sei} which mentions related issues.}, except that a 
geometric interpretation of the nonlocal
transform in \eq{transform} is not available. Hopefully we will
be able to report on this on another occasion.

\section{Conclusion}

(a) We presented a complete derivation of absorption cross-section and
Hawking radiation of minimal and fixed scalars in the D-brane picture
from the five-dimensional black holes starting from the moduli space
of the D1-D5 brane system. In this we have deduced specific CFT
operators coupling to these bulk modes by demanding that the coupling
should respect the symmetries of the theory. These symmetries in the
limit of large $gQ$ should include the symmetries of the near-horizon
geometry, as emphasized by the AdS/CFT correspondence. In addition to
finding the right CFT operator, we construct a gauge-invariant
microcanonical ensemble in the Hilbert space of the orbifold CFT and
calculate gauge-invariant $S$-matrix elements which agree with the
semiclassical result. 

(b) We determine the normalization of the interaction Lagrangian which
couples CFT operators to bulk modes by using the quantitative version
of the AdS/CFT correspondence, where we compare two-point functions
computed from CFT and from supergravity around AdS$_3$ background.
The normalization fixed this way remarkably leads to precise equality of
absorption cross-sections (consequently Hawking radiation rates)
computed from CFT and from supergravity around the black hole
background.

(c) We have computed 2- and 3-point functions of CFT operators
corresponding to minimal scalars (metric fluctuations of the torus)
from tree-level supergravity and by a direct CFT calculation.  We
found that with appropriate choice of normalization of the interaction
Larangian (mentioned above) the two-point functions agree
precisely. The 3-point functions also agree and they both vanish.

(d) We have settled a long-standing problem in the context of fixed
scalars by showing that consistency with near-horizon symmetry demands
that they cannot couple to (1,3) or (3,1) operators. They can only
couple to (2,2) operators. This removes earlier discrepancies between
D-brane calculations and semiclassical calculations of absorption and
emission.

(e) We present a detailed and  explicit classification in Appendix B of
KK modes of IIB supergravity multiplets on $AdS_3 \times S^3 \times
T^4$ in terms of short multiplets of $SU(1,1|2)\times SU(1,1|2)$
(see also \cite{Lar}).

(f) We have given an interpretation of the AdS/CFT (more generally
AdS/SYM) correspondence which relates supergravity equations on
AdS-type backgrounds to large $N$ equations of gauge theory.

(g) We have discussed the explicit nonlocal transform between
the one-dimensional matrix model ($c=1$) and two-dimensional
field theory in the language of holography.

\gap

\noindent {\bf Acknowledgment:} We would like to thank Avinash Dhar
for numerous conversations and Edward Witten for a critical discussion
on instanton moduli spaces. S.W. would like to thank David Gross and
participants of Duality Workshop, Santa Barbara, 1998, and S.W. and
G.M. would like to thank the Organizers and participants of Strings
'98, for providing a stimulating atmosphere in which this work was
begun. We would also like to thank Sumit Das and Fawad Hassan for
some useful comments on the previous version of this paper.

\appendix

\section{The Supergravity Equations}

\gap

We begin with the bosonic sector of Type IIB supergravity. The
Lagrangian is (we follow the conventions of \cite{BACHAS}) 
\bea  
I &=& I_{\rm NS} + I_{\rm RR}\nonumber\\
I_{\rm NS} &=& 
-\frac{1}{2k_{10}^2} \int d^{10} x \sqrt{-G}\left[ e^{-2\phi} \left( R -
4(d\phi)^2 + \frac{1}{12} (dB)^2 \right)  \right] \nonumber\\
I_{\rm RR} &=&  -\frac{1}{2k_{10}^2} \int d^{10} x \sqrt{-G}  
\left (\sum_{n=3, 7}\frac{1}{2n!}
(H^{n})^2 \right)
\eea 
We use $\hat{M}, \hat{N} \ldots $ to denote $10$ dimensional indices,
$i,j, \ldots$ to denote coordinates on the torus $T^4$, $M ,N
\ldots $ to denote the remaining $6$ dimensions and $\mu, \nu,
\ldots $ to denote coordinates on the AdS$_3$.  $k_{10}^2= 
64\pi^7 g^2$
(we use $\alpha'=1$). We have separately indicated the terms
depending on NS-NS and RR backgrounds.

Our aim will  be to obtain the Lagrangian of the minimally coupled
scalars in the D1-D5-brane system. We will find the Lagrangian up to
cubic order in the near horizon limit. Let us first focus
on $I_{\rm NS}$. 

The solution of the supergravity equations for the D1-D5 system
in the string metric is the following (see, e.g., \cite{MAL} whose
notations are used below)
\bea
ds^2 &=& f_1^{-\frac{1}{2}} f_5^{-\frac{1}{2}} (-dt^2 + dx_5^2
) + f_1^{\frac{1}{2}} f_5^{\frac{1}{2}} (dx_1^2 + \cdots + dx_4^2) 
\\ \nonumber 
    & & + f_1^{\frac{1}{2}} f_5^{-\frac{1}{2}} 
(dx_6^2 + \cdots + dx_9^2),
\\ \nonumber
e^{-2(\phi_{10} -\phi_{\infty})} &=& f_5 f_1^{-1} , \\ \nonumber
B^{'}_{05} &=& \frac{1}{2} (f_1^{-1} -1), \\ \nonumber
H^{'}_{abc} &=& (dB^{'})_{abc}
=\frac{1}{2}\epsilon_{abcd}\partial_{d} f_5, \;\;\;\;
 a, b, c, d = 1, 2, 3, 4 
\eea
Where we have substituted $N=0$ in the solutions given in
\cite{MAL} and $f_1$ and $f_2$ are given by
\be
f_1 =1 + \frac{c_1 Q_1}{r^2} ,  \;\;\;\; 
f_5 = 1 + \frac{c_5 Q_5}{r^2}
\ee
Here $r^2 = x_1^2 + x_2^2 + x_3^2 + x_4^2,
c_1=16\pi^4 g/V_4, c_5=g$.
We now substitute the above values of the fields in the the Type
IIB Lagrangian with the  following change in the metric
\be
\label{changemetric}
f_1^{\frac{1}{2}} f_5^{-\frac{1}{2}}\delta_{ij} \rightarrow 
f_1^{\frac{1}{2}} f_5^{-\frac{1}{2}}(\delta_{ij} + h_{ij}).
\ee
Where $h_{ij}$ are the minimally coupled scalars. Their trace is
zero. These scalars are functions of the 6 dimensional
coordinates. The Lagrangian unto cubic order in $h$ ignoring the
traces is
\be
I_{\rm NS}= -\frac{V_4}{2k_{10}^2}\int d^6x \sqrt{-G}
\frac{G^{M N}}{4} \left[ \partial_M h_{ij} \partial_N h_{ij} +
\partial_M (h_{ik}h_{kj} ) \partial_N h_{ij} \right]
\ee
In the above equation we have used the near horizon limit 
and $V_4$ is the volume of the $T^4$. The
metric $G_{MN}$ near the horizon is 
\be
ds^2 = \frac{r^2}{R^2} ( -dx_0^2 + dx_5
^2 ) + \frac{R^2}{r^2} dr^2 + R^2 d\Omega_3 ^2
\ee
We make a change of variables to the Poincare coordinates 
by substituting
\bea
z_0 &=& \frac{R}{r} \\  \nonumber
z_1 &=& \frac{x_0}{R}  \\   \nonumber
z_2 &=& \frac{x_5}{R}  
\eea
The metric becomes 
\be
ds^2 = R^2 \frac{1}{z_0^2} ( dz_0^2 - dz_1^2 + dz_2^2) +
R^2 d\Omega_3 ^2 .
\ee
Here $R= (c_1 Q_1 c_5  Q_5)^{1/4}$ is the radius of curvature
of $AdS_3$ (also of the $S^3$).
For $s$-waves the minimal scalars do not
depend on the coordinates of the $S^3$. Combining all
this, $I_{\rm NS}$ accurate till the cubic order in the $h$'s,
is given by 
\be
I_{\rm NS}= -\frac{V_4}{8k_{10}^2} R^3 
V_{S^3} \int d^3z \sqrt{-g} g^{\mu\nu} \left[ \partial_\mu
h_{ij} \partial_\nu h_{ij} + \partial_\mu (h_{ik} h_{kj})
\partial_\nu h_{ij} \right],
\ee

We would now like to show that up to cubic order $I_{RR}=0$.
The relevant terms in our case are 
\be
I_{RR}=-\frac{1}{4\times 3! k_{10}^2} \int d^{10} x \sqrt{-G} 
H_{\hat{M} \hat{N} \hat{O} } H^{\hat{M} \hat{N} \hat{O}}.
\ee
We substitute the values of $B'$ due to the magnetic and electric
components of the RR charges and the value of $G$. The
contribution from the electric part of $B'$, after
going to the  near-horizon
limit and performing the integral over, is 
\be
\frac{V_4}{4 k_{10}^2} R V_{S^3} \int d^3 z \sqrt{-g}
\sqrt{{\mbox{det}}(\delta_{ij} + h_{ij})}
\ee
The contribution of the magnetic part of $B'$ in the
same limit is 
\be
-\frac{V_4}{4 k_{10}^2} R V_{S^3} \int d^3 z \sqrt{-g}
\sqrt{{\mbox{det}}(\delta_{ij} + h_{ij})}
\ee
We note that the contribution of the electric and the magnetic
parts cancel giving no couplings for the minimal scalars
to the RR background.

\section{The Supergravity Spectrum}

In this section we analyze the spectrum of Type IIB string theory
compactified on $AdS_3 \times S^3 \times T^4$. We ignore the KK modes
on the $T^4$. We show that the KK spectrum of the six dimensional
theory on $AdS_3\times S^3$ can be completely organized as short
multiplets of the supergroup $SU(1,1|2)\times SU(1,1|2)$.  We will
follow the method developed by \cite{deB}. 

The massless spectrum of Type IIB on $T^4\times R^{(5,1)}$ consists
of:

a graviton; 8 gravitinos; 5 two forms; 
16 gauge fields; 40 fermions; and 25 scalars.

Since these are massless, they fall into various representations $R_4$
of the little group $SO(4)$ of $R^{(5,1)}$. On further compactifying
$R^{(5,1)}$ into $AdS_3 \times S^3$, each representation $R_4$
decomposes into various representations $R_3$ of $SO(3)$, the local
Lorentz group of the $S^3$.  The dependence of each of these fields on
the angles of $S^3$ leads to decomposition in terms of KK
modes on the $S^3$ which transform according to some representation of
the isometry group $SO(4)$ of $S^3$. Only those representations of
$SO(4)$ occur in these decompositions which contain the representation
$R_3$ of $S^3$. Once the complete set of KK modes are obtained we
organize them into short multiplets of the supergroup $SU(1,1|2)\times
SU(1,1|2)$.

The graviton transforms as $(\re{3},\re{3})$  of the little group
in 6 dimensions. The KK harmonics of this field are
\bea
\label{eq.gravi}
&(\re{1}, \re{1} ) + 2 (\re{2},\re{2}) + (\re{3},\re{1}) +
(\re{1},\re{3})  \\ \nonumber
&+ 3 \oplus_{\re{m}\geq \re{3}} (\re{m},\re{m}) + 2
\oplus_{\re{m}\geq \re{2}} [\, (\re{m} + \re 2,\re{m} ) + 
(\re{m} , \re{m} + \re2 ) \,]  \\ \nonumber
&+ \oplus_{\re{m}\geq \re 1} [\, (\re{m} + \re 4 ,\re{m} ) +
(\re{m}, \re{m} + \re 4 )  \, ]
\eea
The little group representations of the 8 gravitinos is 
$4(\re 2 , \re 3) + 4(\re 3 ,\re 2) $. Their KK 
harmonics are
\bea
& 8 [\, (\re 1, \re 2 ) + (\re 2 , \re 1) \,] + 
16 \oplus_{\re m  \geq  \re 2 } 
[\, (\re m + \re 1 , \re m ) + ( \re m , \re m + \re 1 ) \, ]
\\  \nonumber
&+ 8 \oplus_{\re m \geq \re 1 }[ \, ( \re m + \re 3 , \re m ) +
(\re m , \re m + \re 3 ) \, ]
\eea
The KK harmonics of the 5 two-forms transforming
in  $(\re 1 ,\re 3 ) + (\re 3 , \re 1 )$ of the little group are
\be
10 \oplus_{\re m \geq \re 2} (\re m, \re m ) + 10 \oplus_{\re m \geq
\re 1 } [\, (\re m +\re 2 , \re m ) + ( \re m , \re m + \re 2 ) \, ]
\ee
The KK harmonics of the 16 gauge fields, $(\re 2 , \re 2)$
 are 
\be
\label{eq.gauge}
16 (\re 1 , \re 1 ) + 32  \oplus_{\re m \geq \re 2 } (\re m , \re m )
+ 16 \oplus_{\re m \geq \re 1 } [\, (\re m , \re m + \re 2 ) + ( \re
m + \re 2 , \re m ) \, ]
\ee
The 40 fermions $ 20 (\re 2 , \re 1) + 20 ( \re 1 , \re 2 )$ give
rise to the following harmonics
\be
40 \oplus_{ \re m \geq \re 1 } [\, ( \re m , \re m + \re 1 ) + (
\re m + \re 1 , \re m ) \, ]
\ee 
The 25 scalars $( \re 1 , \re 1 )$  give rise to the harmonics
\be
25 \oplus_{\re m \geq \re 1} (\re m , \re m )
\ee
Putting all this together the complete KK specturm of Type IIB on
$AdS_3 \times S^3 \times T^4$ yields 
\bea
\label{eq.spec}
&42 ( \re 1 , \re 1 ) + 69 (\re 2 , \re 2 ) + 48 [\, (\re 1 , \re
2 ) + ( \re 2 , \re 1 ) \, ] + 27 [ \, (\re 1 , \re 3 ) + ( \re 3
, \re 1 ) \, ]  \\    \nonumber
&70 \oplus_{\re m \geq \re 3 } ( \re m , \re m ) + 56 \oplus_{\re
m \geq \re 2 } [\, (\re m , \re m + \re 1  ) + ( \re m + \re 1 ,
\re m ) \,] \\ \nonumber
&+ 28 \oplus_{\re m \geq \re 2 } [\, ( \re m , \re m + \re 2 ) +
( \re m + \re 2 , \re m ) \, ] + 8 \oplus_{\re m \geq \re 1 } [\,
(\re m , \re m + \re 3 ) + (\re m + \re 3 , \re m ) \, ] \\
\nonumber
&+ \oplus_{\re m \geq \re 1} [ \, ( \re m , \re m + \re 4 ) + ( \re m
+ \re 4 , \re m ) \, ]
\eea
We now organize  the above KK modes into 
short representations of $SU (1,1 | 2) \times SU ( 1,1 |2 ) $
\cite{deB}. The short multiplet of $ SU(1,1 | 2)$ consists of
the following states 
\be
\begin{array}{ccc}
  j      & L_0   & \mbox{Degeneracy}     \\    
\hline 
\vspace{-.2ex}
h     & h       & 2h +1    \\    
h-1/2~~   & ~~h + 1/2 & 2(2h)     \\    
h-1   &h+1 & 2h -1  
\end{array}
\ee 
In the above table $j$ labels the representation of $SU(2)$ which
is identified as one of the $SU(2)$'s of the isometry
group of $S^3$. $L_0$ denotes the conformal weight of the state. 
We denote the short multiplet of $SU(1,1 |2) \times SU(1,1 |2)$ as 
$(\re{2h} + \re{1} , \re{2h'} + \re 1 )_S$. On carefully organizing the
KK spectrum into short multiplets we get the following
set
\bea
\label{eq.short}
&5 (\re 2 , \re 2 )_S + 6 \oplus_{\re m \geq \re 3 } (\re m , \re m
)_S \\ \nonumber &\oplus_{\re m \geq \re 2 } [\, (\re m , \re m + \re2
)_S + ( \re m + \re2 , \re m )_S + 4 ( \re m , \re m + \re 1 )_S + 4 (
\re m + \re 1 , \re m )_S \, ] 
\eea 
Equation \eq{eq.spec} shows that
there are $42 (\re 1 ,\re 1)$ $SO(4)$ representations in the
supergravity KK spectrum. We know that one of these arises from the
$s$-wave of $g_{55}$ from equation \eq{eq.gravi}.  This is one of the
fixed scalars. $16 (\re 1 ,\re 1)$ comes from the $s$-waves of the
$16$ gauge fields (the components along $x^5$) as seen in equation
\eq{eq.gauge}.  The remaining $25$ comes from the $25$ scalars of the
six dimensional theory. We would like to see where these $42 (\re 1
,\re 1)$ fit in the short multiplets of $SU(1,1|2)\times
SU(1,1|2)$. From equation \eq{eq.short} one can read that $20$ of them
are in the $ 5 (\re 2 , \re 2 )_S$ with $(j=0, L_0 =1 ;\, j=0, L_0 =
1)$. 6 of them are in in $6 (\re 3 ,\re 3 )_S$ with $(j=0, L_0=2;\,
j=0, L_0=2)$. These correspond to the fixed scalars. Finally, the
remaining $16$ of them belong to $4(\re 2 , \re 3 )_S + 4(\re 3 , \re
2 )_S $. $8$ of them have $(j=0, L_0=1 ;\, j=0, L_0=2 )$ and $8$ of
them have $(j=0, L_0=2 ;\, j=0, L_0=1 )$. These scalars can be
recognized as the intermediate scalars. We remark that so far as the
scalars are concerned, the symmetry properties do not depend on
whether we are using periodic or antiperiodic boundary conditions on
the fermions. The reason is that the $j$-values and the $L_0$ values
remain the same under spectral flow upto a normal ordering constant
\cite{GivKutSei}.

\end{document}